\documentclass[aps,showkeys,showpacs,tightenlines,nofootinbib,prd]{revtex4}
\usepackage{amsmath,amscd,amssymb}
\usepackage{color,epsfig}
\usepackage{xfrac}
\usepackage{latexsym}
\usepackage{mathrsfs}
\usepackage[caption=false]{subfig}
\usepackage{graphicx}
\usepackage{ulem}
\usepackage{bbm}
\newcommand{\imu}{{\rm i}}

\usepackage[dvipsnames]{xcolor}

\begin{document}

\title{Kink Collision in the Noncanonical $\varphi^{6}$ Model: A Model with Localized Inner Structures.}

\author{I. Takyi$^{a)}$, S. Gyampoh$^{a)}$, B. Barnes$^{a)}$, J. Ackora-Prah$^{a)}$, and G. A. Okyere$^{b)}$}
\affiliation{$^{a)}$Department of Mathematics, Kwame Nkrumah University of Science and Technology, Private Mail Bag, Kumasi, Ghana\\
	$^{b)}$Department of Statistics and Acturial Science, Kwame Nkrumah University of Science and Technology, Private Mail Bag, Kumasi, Ghana\\
Email: ishmael.takyi@knust.edu.gh}

\begin{abstract}
We study collisions of kinks in the one-space and one-time dimensional noncanonical nonintegrable scalar $\phi^{6}$ model. We examine the energy density of the kink, and we find that, as a function of the parameters that control the curvature of the potential, a localized inner structure of the energy density emerges. We also examine the kink excitation spectrum and the dynamics of the kink collisions for a wide range of initial velocities. We find that apart from the resonance windows, the production of two to three oscillons occurs for some values of the principal parameters of the model. 
\end{abstract}
\keywords{Kinks; Scattering theory; Bound states; Domain walls.}
\pacs{03.65.Ge, 05.45.Yv, 11.10.Lm}

\maketitle

\section{Introduction}

In this paper, we consider the collision of kink and antikink in the noncanonical $(1+1)$ space-time dimensional nonintegrable $\varphi^{6}$ model. The collision process for the canonical models shows some richness, as already shown by the authors~\cite{Ablowitz:1979a,Moshir:1981ja,Campbell:1983xu, Belova:1985fg, Anninos:1991un,Goodman:2005ja} in the case of the $\varphi^{4}$ model, where for sufficiently small initial velocities, the kink and antikink capture one another, leading to a bion-state. For larger velocities, the kink and antikink pair reflect after the collision without further interaction. However, for intermediate initial velocities, a sequence of the so-called resonance windows appears between the larger bion region and the one-bounce (reflection) state. In particular are the two-resonance windows, which, according to~\cite{Campbell:1983xu}, are a result of a two-interaction process. In the first interaction, the energy is transferred from the translational zero mode to an internal shape mode oscillation of the kink such that the kink and antikink are unable to overcome their attractive potential. This results in binding them together. In the second interaction, the internal shape mode is destroyed and its energy is turned to the translational mode, and the kink-antikink pair are liberated from their mutual attraction. Moreover, a higher number of resonance windows is observed in conjunction with the higher number of internal mode oscillations present in the model.

These features have also been reported in the literature for several interesting canonical models, for example, the modified sine-Gordon model~\cite{Peyrard:1983rzn} and higher polynomial canonical models such as the $\varphi^{6}$, $\varphi^{8}$ and $\varphi^{10}$ models~\cite{Dorey:2011yw, Weigel:2013kwa, Gani:2014gxa, Gani:2015cda, Marjaneh:2017mko, Belendryasova:2017wad, Christov:2018wsa, Manton:2018deu, Gani:2020wej, Gani:2020pio}. Most of these models have internal shape modes that are responsible for the resonant collisions, except for the $\varphi^{6}$ model, where resonance collisions are reported~\cite{Dorey:2011yw} despite the absence of an internal shape mode. Here, it was found that the resonant collision resulted from a transfer of energy from the translational zero modes to an extended meson state residing in the potential resulting from linear perturbations.

To explain this, the collective coordinate was used~\cite{Takyi:2016tnc} to obtain the kink-antikink attractive potential as a function of the separation of the kink-antikink. This potential is understood intuitively as the energy of the static field configuration. The authors in Ref.~\cite{Demirkaya:2017euk} extended this argument to the symmetric $\varphi^{6}$ model, which was proposed in Ref.~\cite{Christ:1975wt, Lohe:1979mh} as a prototype bag model for studying quarks within hadrons. Usually, the collective coordinate method serves as an analytical approximation~\cite{Christov:2008kk, Manton:1978gf, Manton:2021ipk} and allows one to estimate the force between the kink and antikink.

The canonical model in (1+1) space-time dimensions has been the subject of active research~\cite{Rajaraman:1982is, Manton2004, Vachaspati:2006zz, Vilenkin:2000jqa}. The dynamical properties of these models, coupled with their resonant collision, are widely used to model various physical phenomena: They are used to study ferromagnetic and ferroelectric properties of materials in condensed matter physics~\cite{Ivanov:1992aa, Trullinger:1976aa,Kevre2008}, as well as phenomenological properties of matter in high energy physics~\cite{Greenwood:2008qp, Ahlqvist:2014uha}, domain walls in cosmology~\cite{Anninos:1991un}, and hadron properties in nuclear physics~\cite{Weigel:2008zz, Weigel:2021pbr}. 

The noncanonical theories, normally referred to as $K-$ defects theories, have been of interest recently. One of their remarkable applications is in explaining the universe's accelerated expansion~\cite{Chiba:1999ka}. For example,~\cite{Gomes:2013bca} investigated the kink-antikink collision for $\varphi^4$ twin models, and~\cite{Zhong:2019fub} investigated the collision of kinks in the noncanonical $\varphi^4$ model. In their studies, they found that the model supports kink and antikink solutions with localized inner structures in their energy density, and the presence of these leads to the formation of oscillons. Takyi et al~\cite{Takyi:2021jzx} extended their research to the noncanonical sine-Gordon model, observing bion structures similar to the canonical sine-Gordon model when they interact with defects or impurities. 

The current study investigates how localized inner structures coupled with internal shape modes generate oscillons and resonance structures in the kink and antikink collision of symmetric noncanonical $\varphi^6$ models. Also, a comprehensive analysis of the resonance mechanism is carried out to ascertain quantitatively how the resonance window centers scale with the number of cyclic oscillations. These issues will be the objectives of the present 
paper. As a starting point, we introduce $(1+1)$ space-time dimensional noncanonical scalar $\varphi^{6}$ field theory and discuss their dynamics in section \ref{sec:models}. Section \ref{sec:numerics} specializes in the numerical analysis of the kink and antikink collisions of the model. Finally, we will end the work with a conclusion in section \ref{sec:conclude}.

\section{The Model}
\label{sec:models}
Here we consider the noncanonical $\varphi^{6}$ model in $(1+1)$ dimensions, where the scalar field is coupled to its kinetic and gradient term via the total energy
\begin{equation}
	E = \int_{-\infty}^{\infty} \left[ \frac{1}{2} F(\varphi) \left(\frac{\partial \varphi}{\partial t}\right)^{2} + \frac{1}{2} F(\varphi) \left(\frac{\partial \varphi}{\partial x}\right)^{2} + U(\varphi)  \right]\, \mathrm{d}x, \label{eq:toenerg}
\end{equation}
where $F(\varphi) = \alpha \varphi^{2m} + 1$. The parameter $\alpha$ is assumed to be real and positive, and it distinguishes our model from the canonical case, whereas the positive integer parameter $m$ controls the inner structures in the kink's energy density. At $\varphi(x \rightarrow \infty) > \varphi(x \rightarrow -\infty)$, the function $U(\varphi)$ is the scalar potential with at least two vacua. The kink(antikink) solutions interpolate between these neighboring vacua. 

We construct the kink solution by first considering the static wave equation 
\begin{equation}
	2 \frac{\partial U}{\partial \varphi} = \frac{\partial F}{\partial \varphi} W(\varphi)^{2} + 2F(\varphi) \frac{\partial^{2}\varphi }{\partial x^{2}} \label{eq:staticeq}
\end{equation} 
where 
\begin{equation}
	W(\varphi) = \frac{\partial \varphi}{\partial x} \label{eq:superpot}
\end{equation}
is the superpotential \cite{Zhong:2019fub,Bazeia:2008tj,Zhong:2014kha,Zhong:2018tbi}. Multiplying Eq.\eqref{eq:staticeq} by $W$, integrating over $x$ and taking the integration constant to be zero gives
\begin{equation}
	U(\varphi) = \frac{1}{2} F(\varphi)W(\varphi)^{2}. \label{eq:scalpotdef}
\end{equation}
Clearly, a given superpotential yields a particular kink solution. In this paper, we define the superpotential as
\begin{equation}
	W(\varphi) = \sqrt{\varphi^{2}+1} \left(\varphi^{2}-1\right) \label{eq:givenpot}.
\end{equation}
The specified superpotential Eq.\eqref{eq:givenpot} was first introduced as a simple one-dimensional `bag' model in which the kinks were identified as quarks within hadrons~\cite{Demirkaya:2017euk,Christ:1975wt,Lohe:1979mh}. Here we consider the simplest form of this model, which is symmetric with two degenerate vacua given by $\varphi_{\rm vac}=\pm 1$. The static equation cf. Eq.\eqref{eq:staticeq} allows for static soliton solutions that connect neighboring vacuum solutions at spatial infinity. In this model, these are the kink and antikink solutions, which upon using Eq.\eqref{eq:superpot} is
\begin{equation}
	\varphi_{K(\bar{K})}(x) = \pm \frac{\sqrt{2}\sinh \left(\frac{\mu x}{2}\right) }{\sqrt{2 + 2 \left(1+\sinh\left(\frac{\mu x}{2}\right)^{2} \right)   }}. \label{eq:kinkprofile}
\end{equation}
Here $\mu=2\sqrt{2}$ denotes the kink fluctuation mass.  Thus, $\mu$ is the strength of the restoring force. The scalar potential on using Eq. \eqref{eq:scalpotdef} then becomes
\begin{equation}
	U(\varphi) = \frac{1}{2} \left(\varphi^{2}+1\right) \left(\varphi^{2}-1\right)^{2} \left(\alpha\varphi^{2m}+1\right)\label{eq:scalarpot}.
\end{equation}  
\begin{figure}
	\centering
	\subfloat[Scalar potential $U(\varphi)$ cf. equation \eqref{eq:scalarpot}.]
	{\includegraphics[scale=0.27]{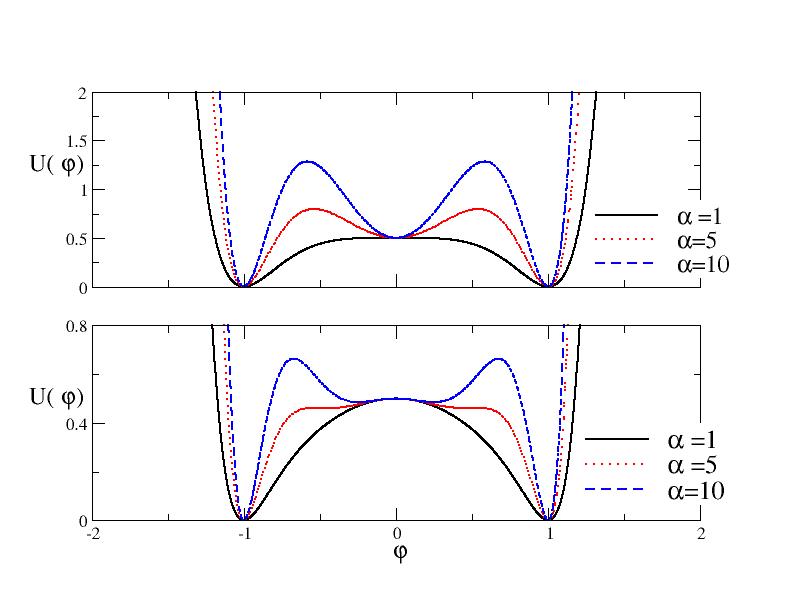}
		\label{fig:scalarpot_a}}
	\quad 
	\subfloat[Energy density of the noncanonical model.]
	{\includegraphics[scale=0.27]{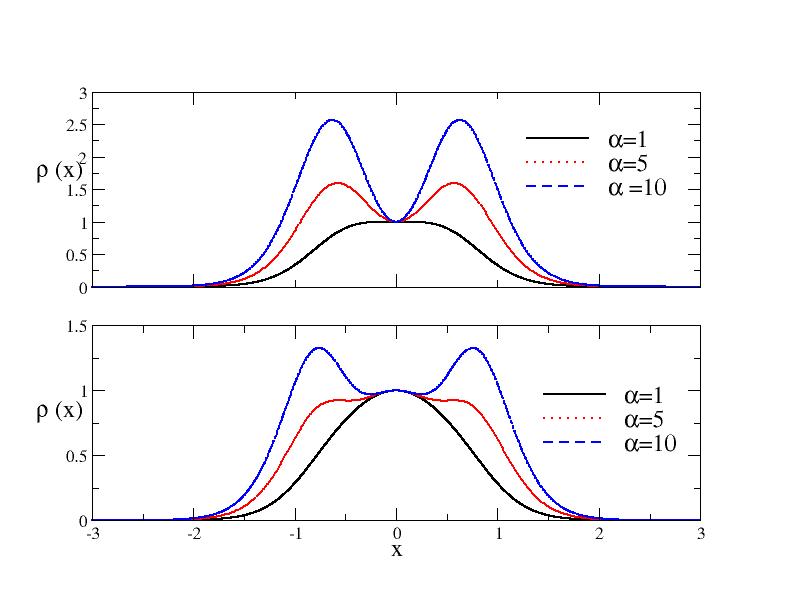} 
		\label{fig:scalarpot_b}}
	\caption{\label{fig:scalarpot} The field potential $U(\varphi)$ and the energy density $\rho (x)$ for $m=1$(top row) and $m=2$(bottom row).}
\end{figure}
We observe from figure \ref{fig:scalarpot_a} that $\alpha$ is the key parameter that controls the curvature of the potential and is critical to understanding the spectral and dynamical structures of our model. The scalar potential transits from a double to a triple well as $\alpha$ varies, with a local minimum at $\varphi_{\mathrm{vac}}=0$ in the case of $m=1$ and a local maximum at $\varphi_{\mathrm{vac}}=0$ in the case of $m=2$. Figure \ref{fig:scalarpot_b} is the energy density, $\rho (x) = 2 U(\varphi_{K})$ for various values of $\alpha$. We observe from here two to three localized inner structures of the energy density for $m=1$ and $m=2$ as the parameter $\alpha$ varies. Substituting the kink profile cf. Eq. \eqref{eq:kinkprofile} into Eq. \eqref{eq:toenerg} gives the classical energy of the kink 
\begin{equation}
	E_{\rm{cl}} = \frac{\sqrt{2}}{4} + \frac{5}{8}\ln \left(\frac{\sqrt{2} +1}{\sqrt{2}-1}\right) + \frac{25\alpha}{\left(20m(m+2) + 16\right)}. \label{eq:Emass}
\end{equation}

In what follows, we analyze the excitation spectrum of the static kink $\varphi_{K}(x)$ by adding to it a small perturbation
\begin{equation}
	\varphi(t,x) = \varphi_{K}(x) + \delta \varphi(t,x)
\end{equation}
and taking in the equation of motion, we obtain terms linear in $\delta \varphi(t,x)$:
\begin{align}
	\frac{\partial^{2}F}{\partial \varphi_{K}^{2}}\left(\frac{\partial \varphi_{K}}{\partial x}\right)^{2} \delta \varphi + 2\frac{\partial F}{\partial \varphi_{K}} \frac{\partial \varphi_{K}}{\partial x} \frac{\partial \delta \varphi}{\partial x}  + 2 F\left( \frac{\partial ^{2} \delta \varphi }{\partial x^{2}} - \frac{\partial^{2} \delta \varphi}{\partial t^{2}} \right) + 2\frac{\partial F}{\partial \varphi_{K}} \frac{\partial^{2} \varphi_{K}}{\partial x^{2}} \delta \varphi - 2\left( \frac{\partial^{2} U}{\partial \varphi_{K}^{2}} + 2 \frac{\partial F}{\partial \varphi_{K}} \frac{\partial W}{\partial \varphi_{K}}\right) \delta\varphi =0.\label{eq:scattering}
\end{align}
Writing $\delta \varphi$ in the form~\cite{Takyi:2021jzx}
\begin{equation}
	\delta\varphi(t,x) = \frac{1}{\sqrt{F}} \psi(x) e^{\imu \omega t}
\end{equation}
we obtain from Eq. \eqref{eq:scattering} the schr\"odinger equation 
\begin{equation}
	\left[-\frac{\mathrm{d}^{2}}{\mathrm{d}x^{2}} + \theta(x)\right] \psi(x) = \omega^{2} \psi(x)
	\label{eq:wave_equ}
\end{equation}
where
\begin{equation}
	\theta(x) = \frac{1}{F} \frac{\partial^{2} U}{\partial \varphi_{K}^{2} } + \frac{3}{2} \frac{1}{F} \frac{\partial F}{\partial \varphi_{K}} \frac{\partial^{2} \varphi_{K}}{\partial x^{2}} - \frac{1}{4} \left(\frac{1}{F} \frac{\partial F}{\partial \varphi_{K}} \frac{\partial \varphi_{K}}{\partial x} \right)^{2} \label{eq:effective_pot}
\end{equation}
is the effective potential. Its explicit expression can easily be obtained by substituting the kink profile. 
\begin{figure}
	\centering 
	\subfloat[The effective potential cf. Eq. \eqref{eq:effective_pot} for $m=1$ (top) and $m=2$ (down).]{\includegraphics[scale=0.27]{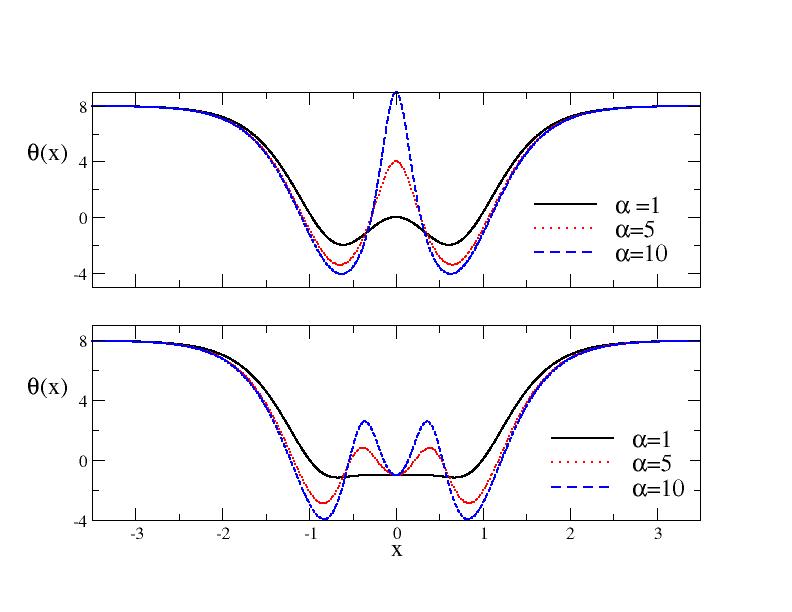} \label{fig:effective}}
	\quad
	\subfloat[The eigen modes $(\omega_{0}, \omega_{1}, \omega_{2}, \omega_{3})$.]{\includegraphics[scale=0.27]{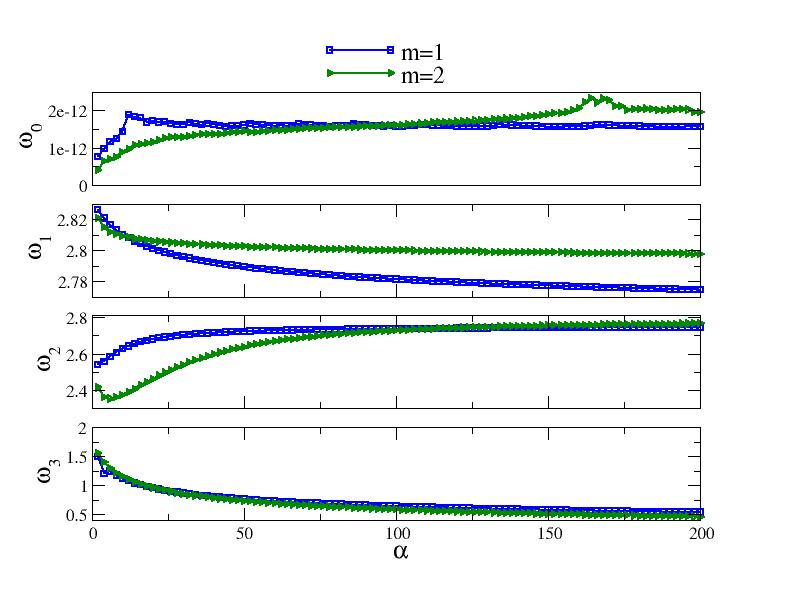}
		\label{fig:eigenvalue}}
	\caption{\label{fig:eigenvalueef} The effective potential and eigen modes of equation \eqref{eq:wave_equ} for $\alpha \in \left[0:2:200 \right]$ and $m=1,2$.}
\end{figure}
Figure \ref{fig:effective} shows the shape of the effective potential. When $\alpha$ increases for $m = 1$, the point $x = 0$ becomes a local maximum with two minima in the potential, whereas when $\alpha$ increases for $m = 2$, the point $x = 0$ becomes a local minimum with three minima in the potential. Furthermore, the effective potential is symmetric, with the asymptotic value $\theta(x \rightarrow \pm \infty) = \mu^2$. We numerically solve the eigenvalue problem of Eq. \eqref{eq:wave_equ} and plot the eigenvalues in figure \ref{fig:eigenvalue} for $m=1,2$ and $\alpha\in\left[0:2:200\right]$. We found from our numerical results that, apart from the zero mode frequency $\omega_{0}$, three additional shape modes are found, $\omega_{1}$, $\omega_{2}$ and $\omega_{3}$. As $\alpha$ varies, $\omega_{1}$ and $\omega_{3}$ decreases, while $\omega_{2}$ increases monotonically.

\section{Numerical Results}
\label{sec:numerics}

\begin{figure}
	\centering 
	\includegraphics[scale=0.3]{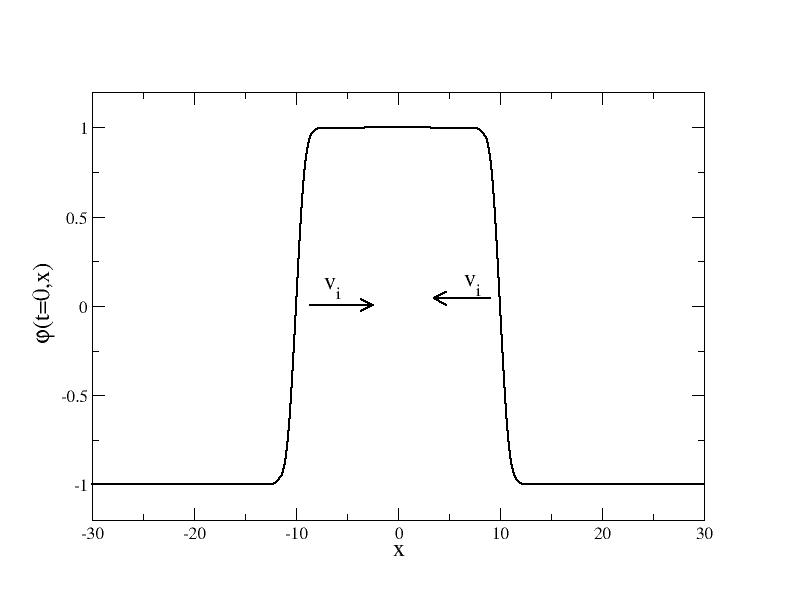} 
	\caption{\label{fig:initialcond} Initial configuration for the kink-antikink collision.}
\end{figure}

In this section, we present the results from the numerical simulations of the kink-antikink collision of the noncanonical $\phi^{6}$ model. This is done by solving the dynamical equation 
\begin{equation}
	2F \frac{\partial^{2} \varphi}{\partial t^{2}} = 2F \frac{\partial^{2} \varphi}{\partial x^{2}} + \frac{\partial F}{\partial \varphi} \left[ \left(\frac{\partial \varphi}{\partial x}\right)^{2} - \left(\frac{\partial \varphi}{\partial t}\right)^{2}\right] - 2 \frac{\partial U}{\partial \varphi}
\end{equation}
subject to the initial condition 
\begin{equation}
	\varphi(t=0,x) = \varphi_{\bar{K}} \left(\frac{x-x_{0}}{\sqrt{1-v_{i}^{2}}}\right) + \varphi_{K} \left(\frac{x+x_{0}}{\sqrt{1-v_{i}^{2}}}\right)-1. \label{eq:spec}
\end{equation}
Here, $\varphi_{K}$ and $\varphi_{\bar{K}}$ are the static kink and the static antikink cf. Eq. \eqref{eq:kinkprofile}. The above ansatz corresponds to the kink and the antikink, separated at $t=0$ by $2x_{0}$, and propagating towards each other with the velocities $\pm v_{i}$ in the rest frame (see figure \ref{fig:initialcond}). In our calculations, we used $2x_{0}=20$, which is larger than the width of the kink according to standard practice. 
With this in place, we then solve the dynamical equation as an initial value problem using the fourth-order center difference scheme to approximate the first and second spatial derivatives, respectively:
\begin{align*}
	\frac{\partial \varphi}{\partial x} & = \frac{1}{12 h} \left(\varphi_{n-2}-8\varphi_{n-1}+8\varphi_{n+1}-\varphi_{n+2}\right) \\
	\frac{\partial^{2} \varphi}{\partial x^{2}} & = \frac{1}{12 h^{2}} \left(-\varphi_{n-2}+16\varphi_{n-1}-30\varphi_{n}+16\varphi_{n+1}-\varphi_{n+2}\right) \\
\end{align*}
where $h$ is the space grid spacing and $n$ number the corresponding grid points coordinates $x_{n}$. In standard practice, a large grid is chosen for the spatial coordinates in such a way that radiation emitted during the collision does not propagate back into the grid after reaching the boundaries. Accordingly, we chose the grid with $8001$ nodes for $x\in \left[-100,100\right]$. The dynamical equation is then propagated in time using adaptive step size control. To check the accuracy of the numerical calculation, we verify that the total energy as in Eq. \eqref{eq:toenerg} is conserved.
\begin{figure}
	\centering
	\subfloat[ Production of three oscillons for $v_{\rm i}=0.295$ for $m=2$ and $\alpha = 5$.]
	{\includegraphics[scale=0.12]{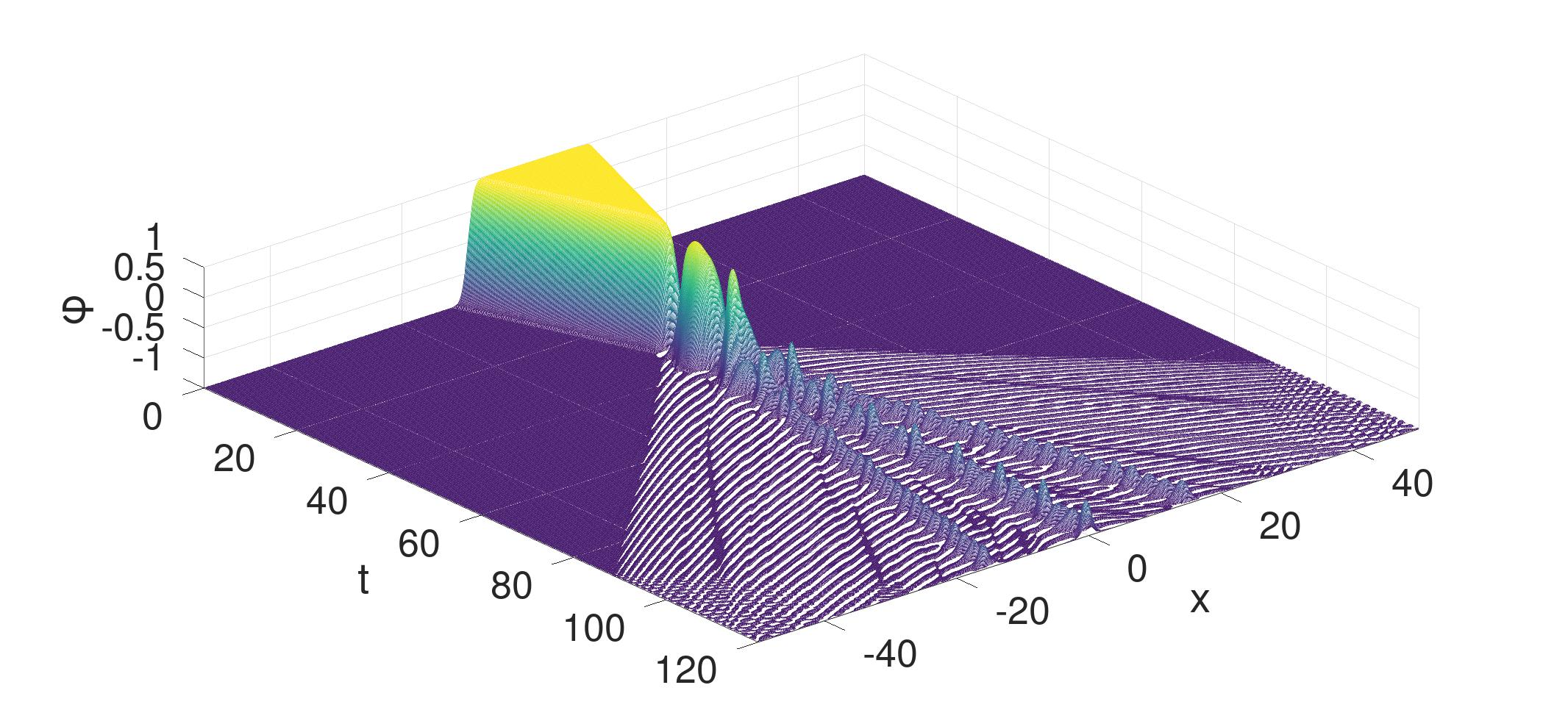} 
		\includegraphics[width=6.8cm,height=4cm]{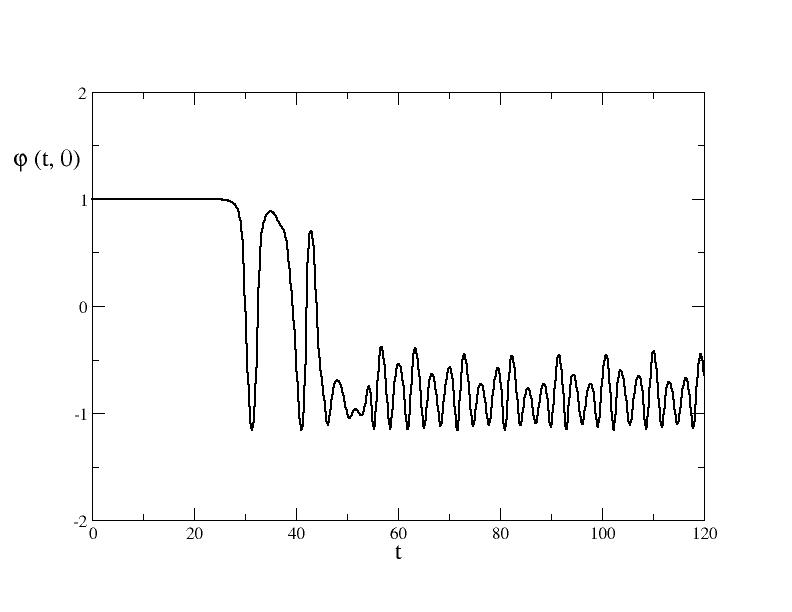} \label{fig:three_oscillon_a}}
	\quad
	\subfloat[ Production of three oscillons for $v_{\rm i}=0.156$ for $m=1$ and $\alpha = 1$.]
	{\includegraphics[scale=0.12]{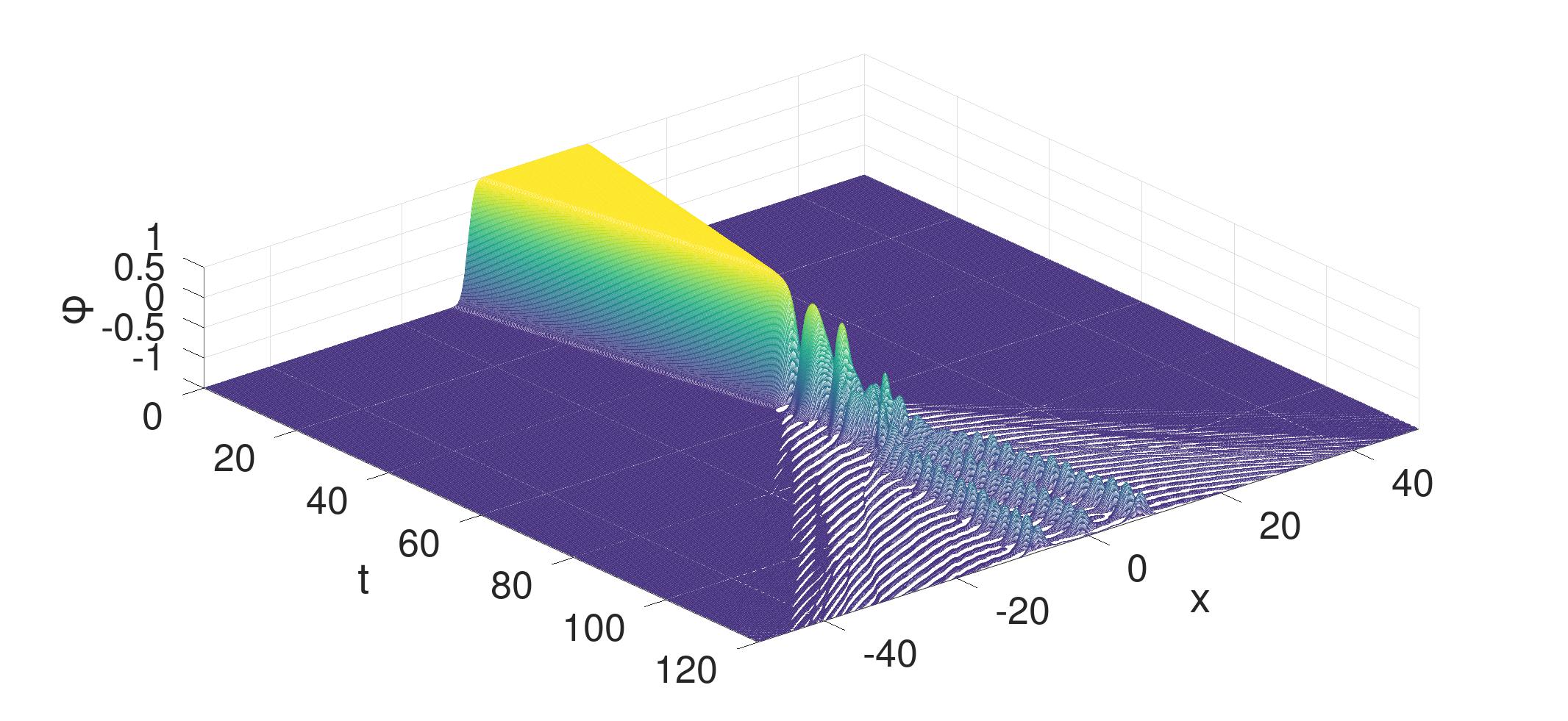}
		\includegraphics[width=6.8cm,height=4cm]{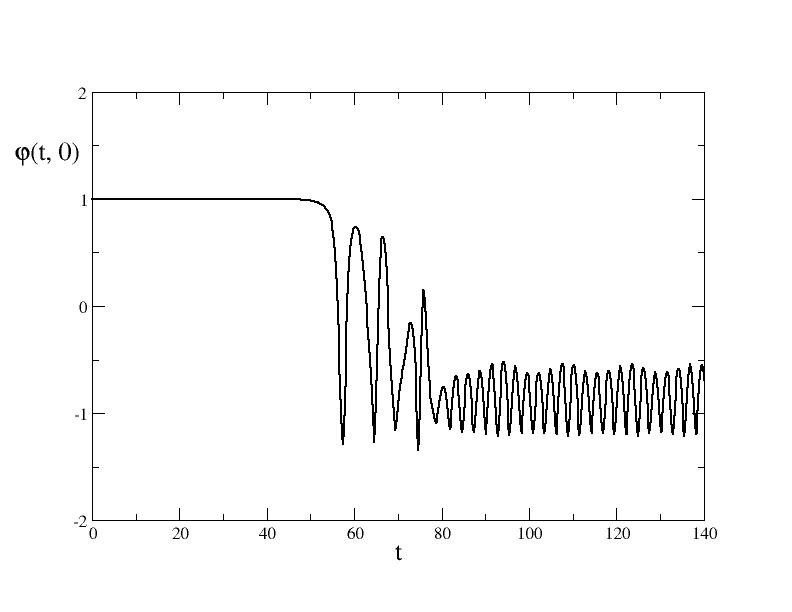} \label{fig:three_oscillon_b}}
	\quad
	\subfloat[Production of two oscillons for $v_{\rm i }=0.271$ for $m=2$ and $\alpha = 5$.]
	{\includegraphics[scale=0.1]{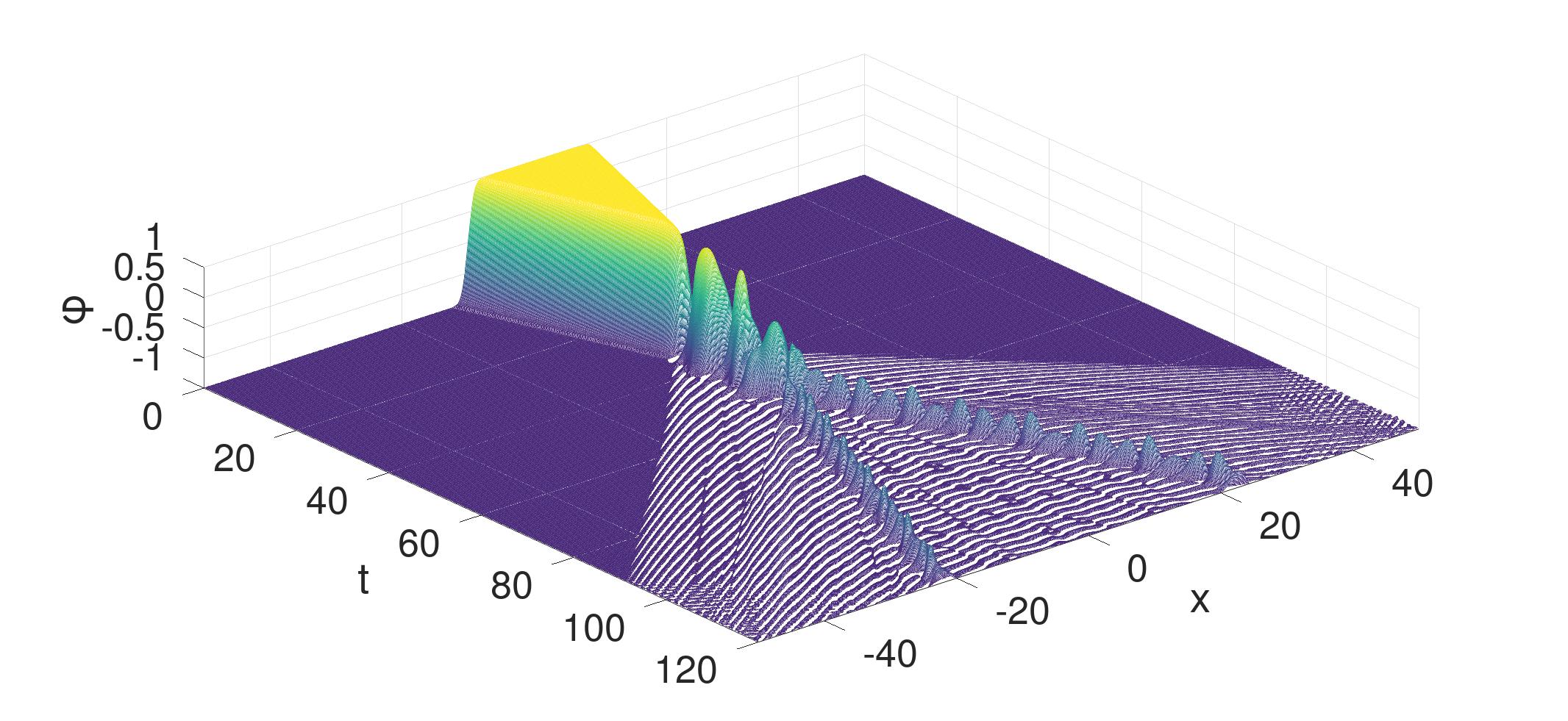}
		\includegraphics[width=6.8cm,height=4cm]{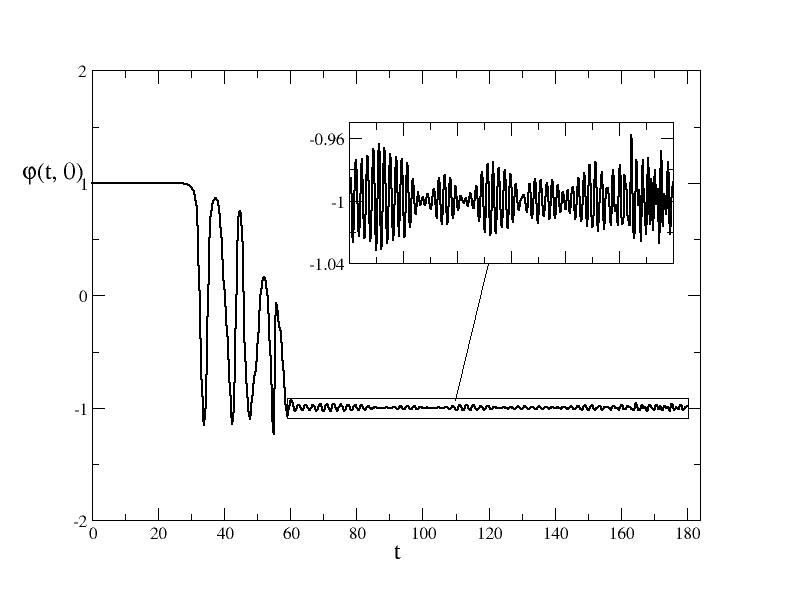} \label{fig:two_oscilon_a}}
	\quad
	\subfloat[Production of two oscillons for $v_{\rm i }=0.182$ for $m=1$ and $\alpha = 1$.]
	{\includegraphics[scale=0.1]{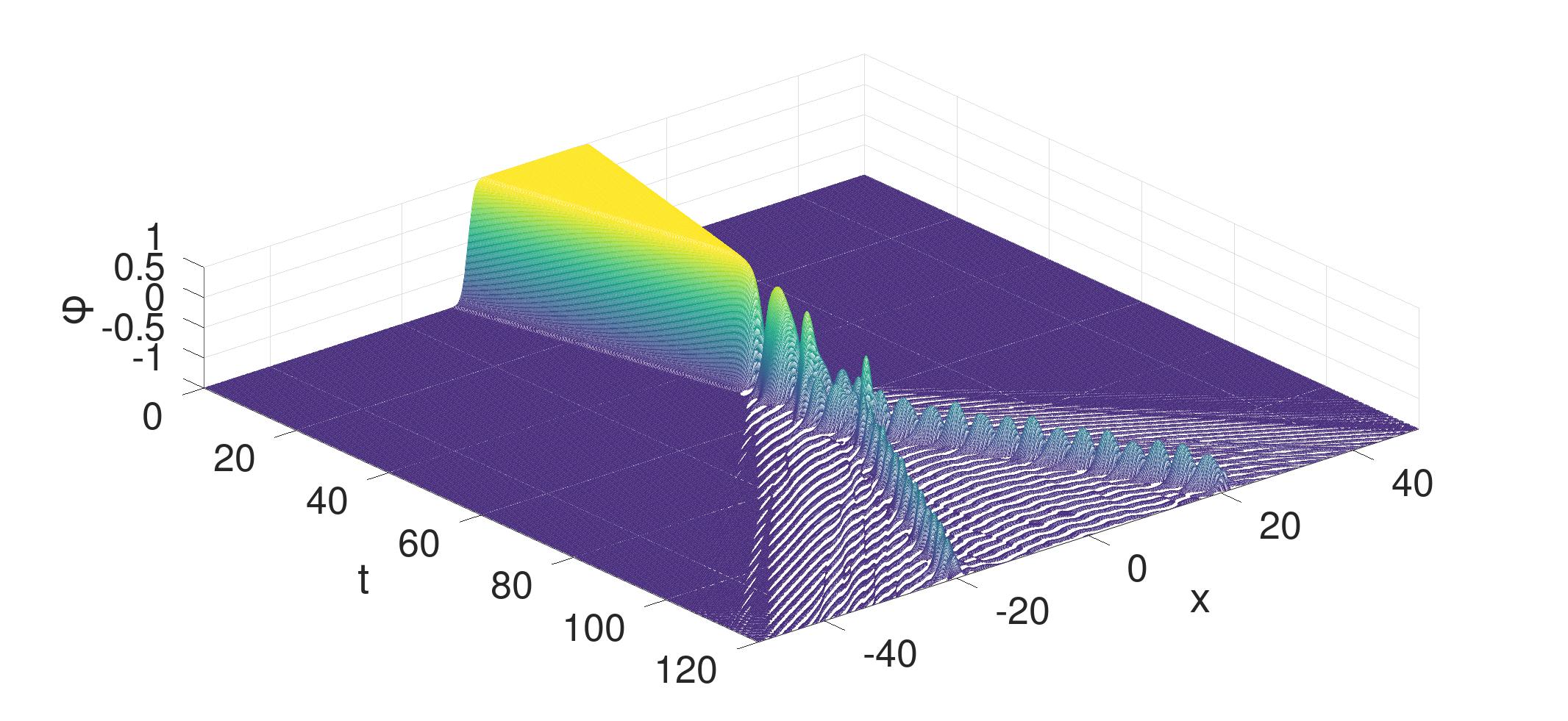}
		\includegraphics[width=6.8cm,height=4cm]{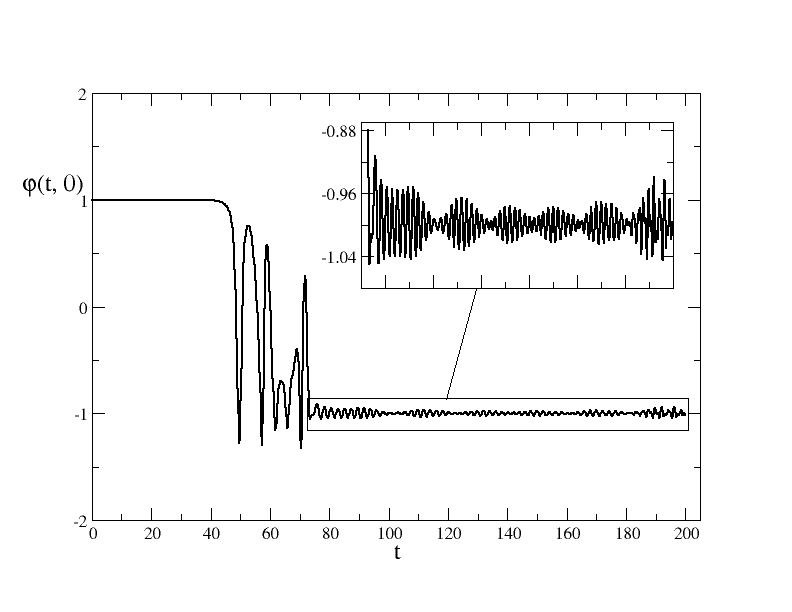}	 \label{fig:two_oscilon_b}}
	\caption{\label{fig:oscillons}Production of oscillons for $v_{i}<v_{c}$. Left panel: the space-time picture. Right panel: time dependence of the field at $x=0$.}
\end{figure}

\begin{table}
	\caption{\label{t3} Predictions for the critical velocities.}
	\centering{
		\begin{tabular}{c|c|c}
			$m$ & $\alpha$ & $v_{\rm c}$ \\
			\hline
			1   &   1      & 0.2490 \\
			1   &   2      & 0.2930 \\
			1   &   5      & 0.4067 \\ 
			1   &   10     & 0.5771 \\ 
			2   &   1      & 0.2093 \\
			2   &   5      & 0.3084 \\
			2   &   10     & 0.3538 \\ 
		\end{tabular}
	}
\end{table}

From our numerical calculations, we observe interesting scattering phenomena depending on the initial propagating velocity $(v_{i})$ and the free parameters that control the inner structures of the kink. For example, the production of two oscillons has been reported and discussed in models such as the noncanonical $\phi^{4}$ model~\cite{Zhong:2019fub}, the double sine-Gordon model~\cite{Campbell:1986nu, Gani:2017yla, Gani:2019jzc}, and the sinh-deformed $\phi^4$ model~\cite{Bazeia:2017rxo,Bazeia:2019nsh} for $v_{i}<v_{c}$. These oscillons are long-lived oscillations (with low amplitude) of the scalar field around one vacuum. These features are shown in figure \ref{fig:oscillons}, where we observed three oscillon productions for $v_{i}=0.295$ ($v_{i}=0.156$) and two oscillon productions for $v_{i}=0.271$ ($v_{i}=0.182$) for $m=2,\alpha=5$ ($m=1,\alpha=1$). The right panel of figure \ref{fig:oscillons} shows the amplitude and frequency of oscillations of these structures. The simulated intrinsic frequency for $m=2, \alpha=5 $ at $v_{i}=0.295$ and $v_{i}=0.271$ is $2.402$ and that for $m=1, \alpha=1$ at $v_i=0.156$ and $v_i=0.182$ is $1.636$. It can be observed in figure \ref{fig:three_oscillon_a} that the central oscillon displays irregular behavior where there is a repeated pattern of a high peak followed by two low peaks compared to figure \ref{fig:three_oscillon_b}. This is attributed to the behavior of the scalar potential in this regime (i.e $m = 2$, $\alpha = 5$). Here the kink has an irregular hill top curvature with a local maximum at $\varphi_{\rm vac}=0$ with a small steep towards large field values. Hill top potentials with larger steepness tend to favor regular oscillons with the same peaks as in the case of double sine-Gordon potentials and sinh-deformed potentials~\cite{Gani:2017yla, Gani:2019jzc,Bazeia:2017rxo}.  
\begin{figure}
	\centering
	\subfloat[Four-resonance window for $m=1, \alpha=2, v_{i}=0.280$.]
	{\includegraphics[scale=0.1]{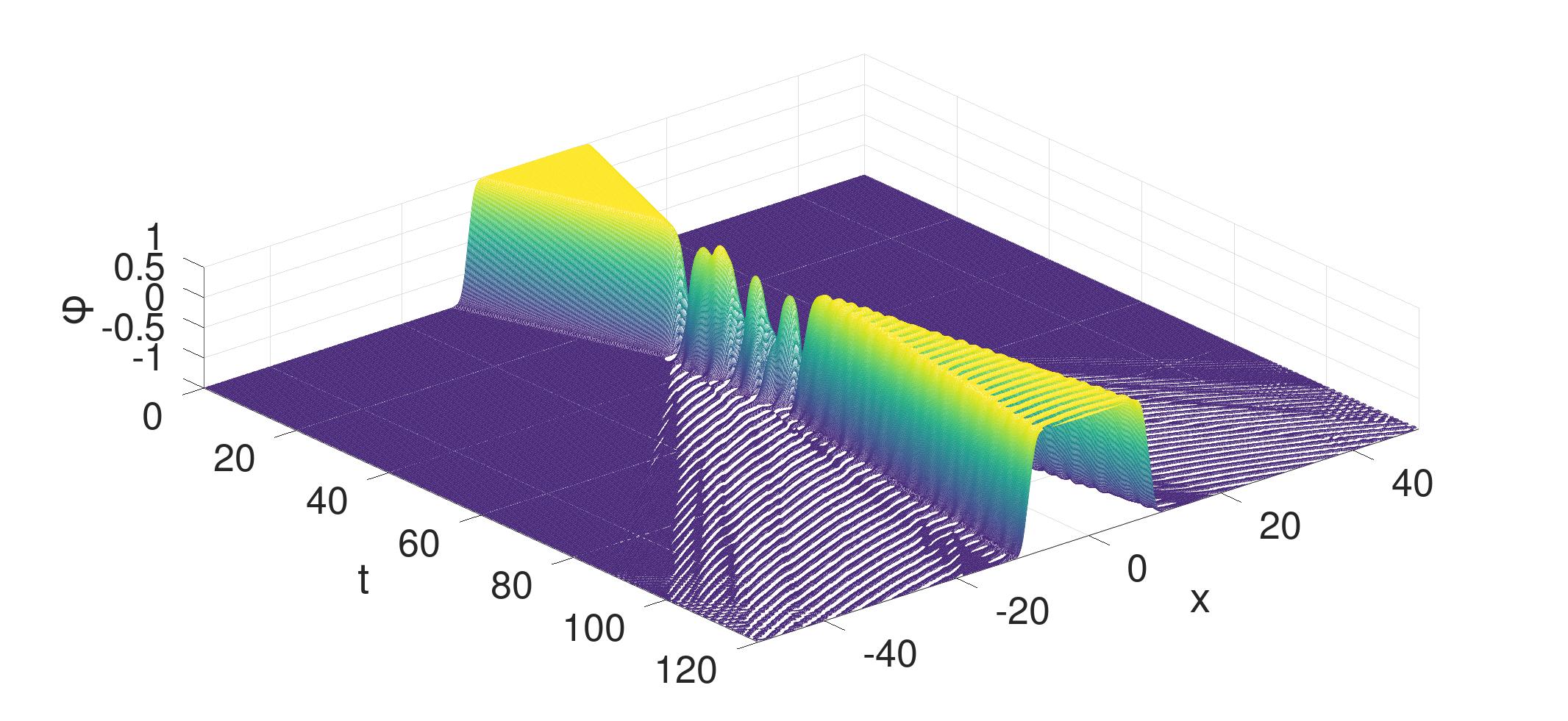}
		\label{fig:four_resonance}}
	\quad 
	\subfloat[Three-resonance window for $m=1, \alpha=1, v_{i}=0.224$.]
	{\includegraphics[scale=0.1]{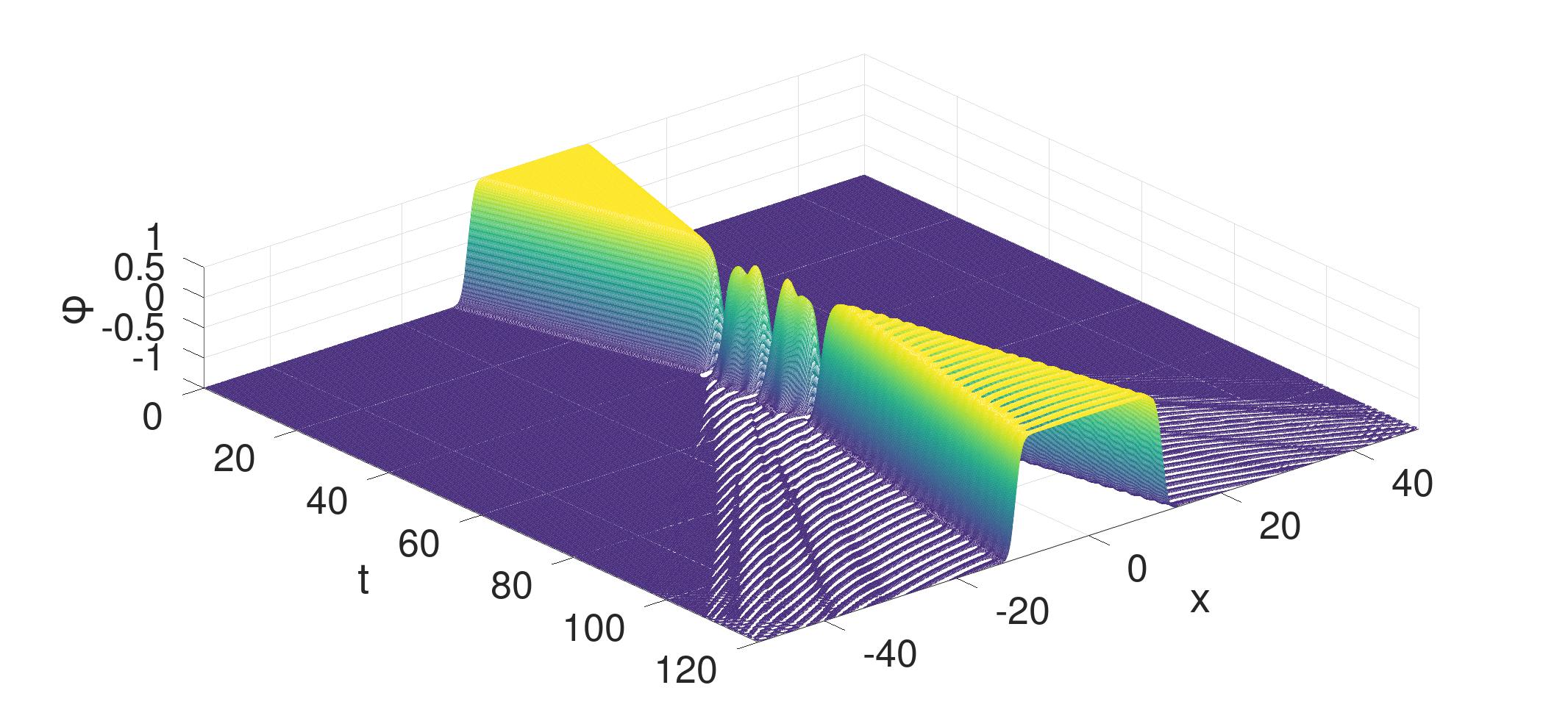} 
		\label{fig:three-resonance}}
	\quad 
	\subfloat[First two-resonance window for $m=2, \alpha=1, v_{i}=0.155$.]
	{\includegraphics[scale=0.1]{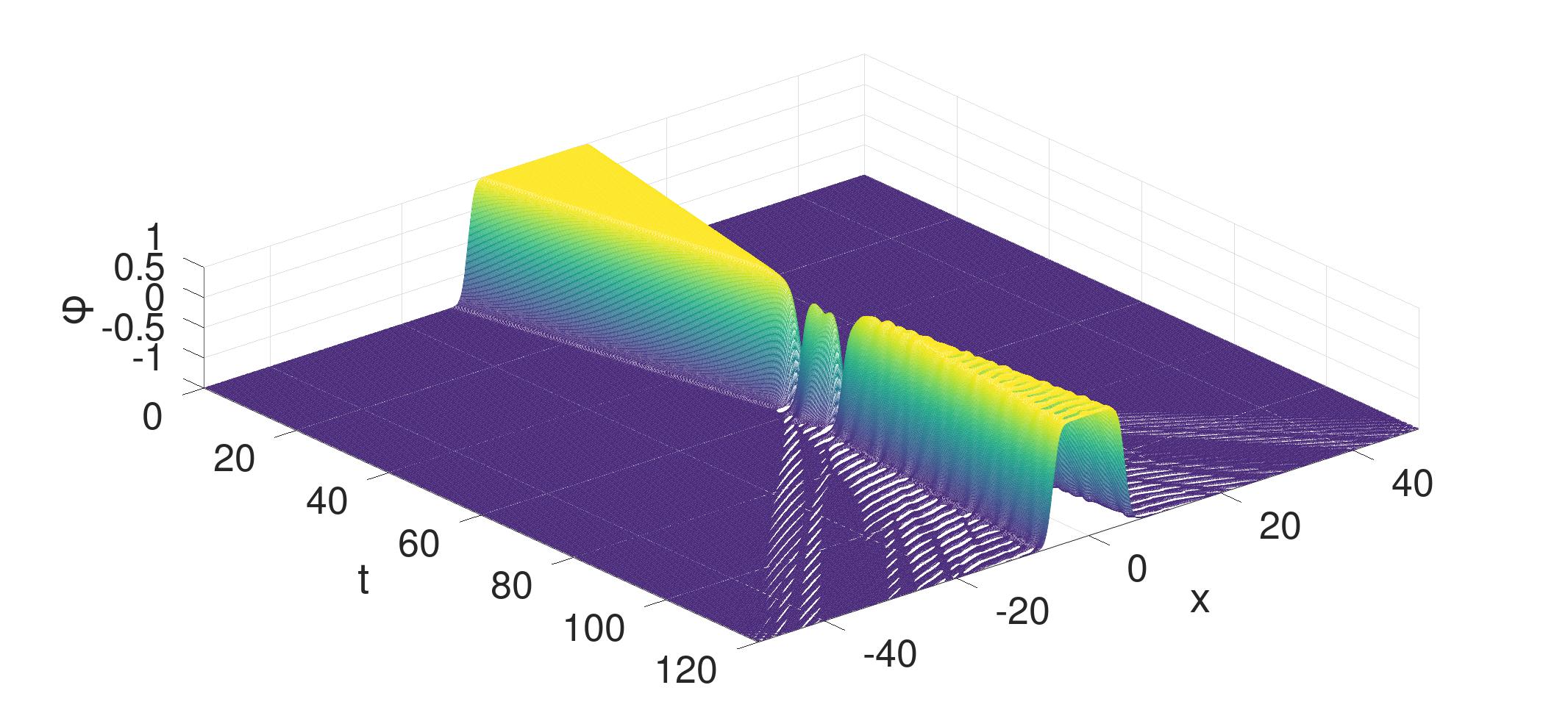} 
		\label{fig:first_two-resonance}}
	\quad 
	\subfloat[Second two-resonance for $m=2, \alpha=1, v_{i}=0.1571$.]
	{\includegraphics[scale=0.1]{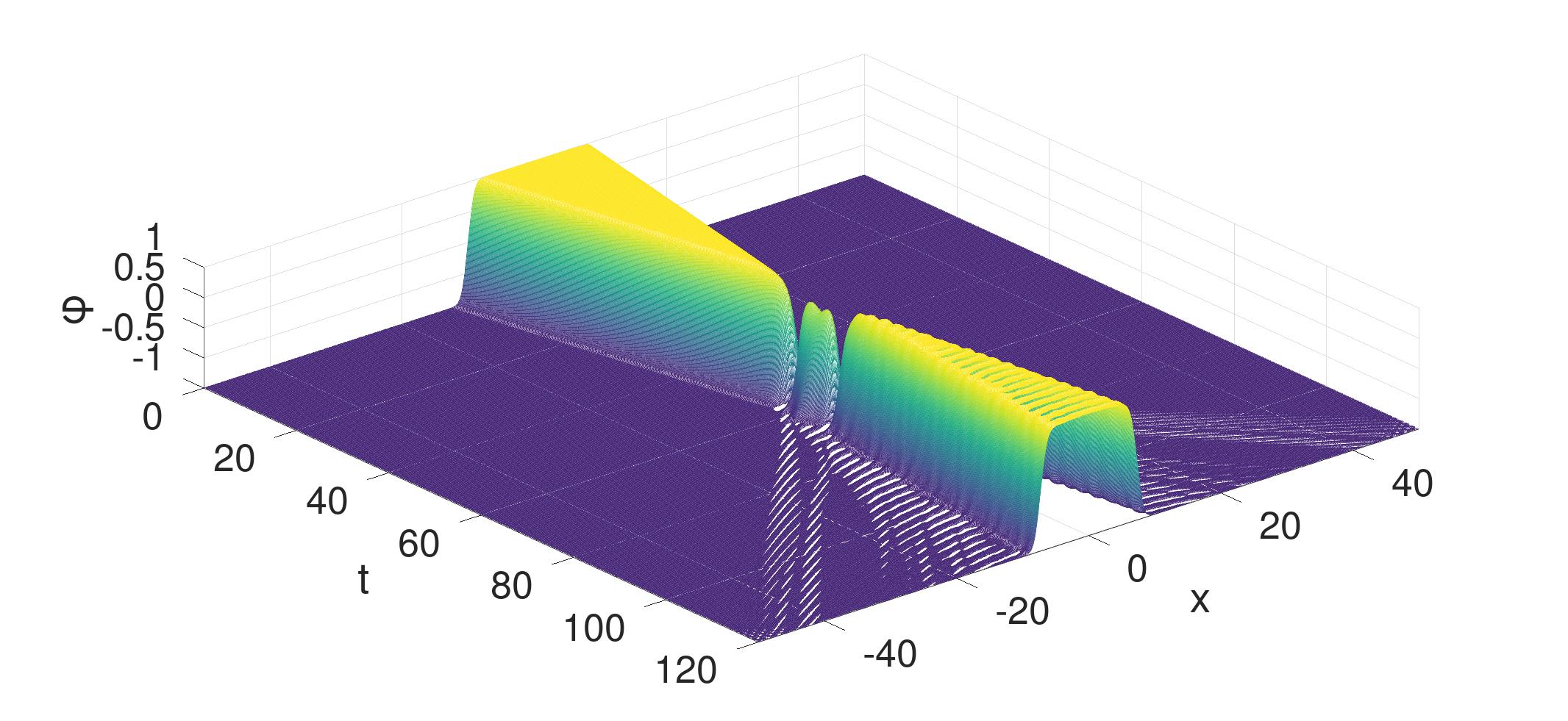} 
		\label{fig:second_two-resonance}}
	\caption{\label{fig:resonace_mechanism} The plot of $\phi(t,x)$ as a function of $v_{i}$ for $m=1$ and $m=2$ showing resonance mechanisms.}
\end{figure}

In other regimes, with initial propagating velocity $v_{i}<v_{c}$, there are intervals of propagating velocities, known as resonance windows, where the kinks escape to spatial infinity after two or more collisions. The observed feature is as a result of the exchange energy between the translational mode and the resonance frequency, `$\omega_{R}$'. The value of $\omega_{R}$ can either coincide with the internal shape modes frequencies of $\omega_i$ of the kink as in the $\phi^4$ model~\cite{Campbell:1983xu, Campbell:1986mg}, or deviate from it as in the double sine-Gordon model~\cite{Campbell:1986nu, Gani:1998jb}. For various values of $m$ and $\alpha$, we observe a number of two-, three-, and four-resonance windows. For example, at an initial propagating velocity of $v_{i} = 0.224$, we observe a three-resonance window for the free parameters $m = 1, \alpha = 1$ and a four-resonance window for $m = 1, \alpha = 2$ at initial velocity of $v_{i}=0.280$. However, for $m=2, \alpha=1$ we observed around eleven two-resonance windows. We illustrate these phenomena in figure \ref{fig:resonace_mechanism}.
\begin{figure}
	\centering
	\includegraphics[scale=0.35]{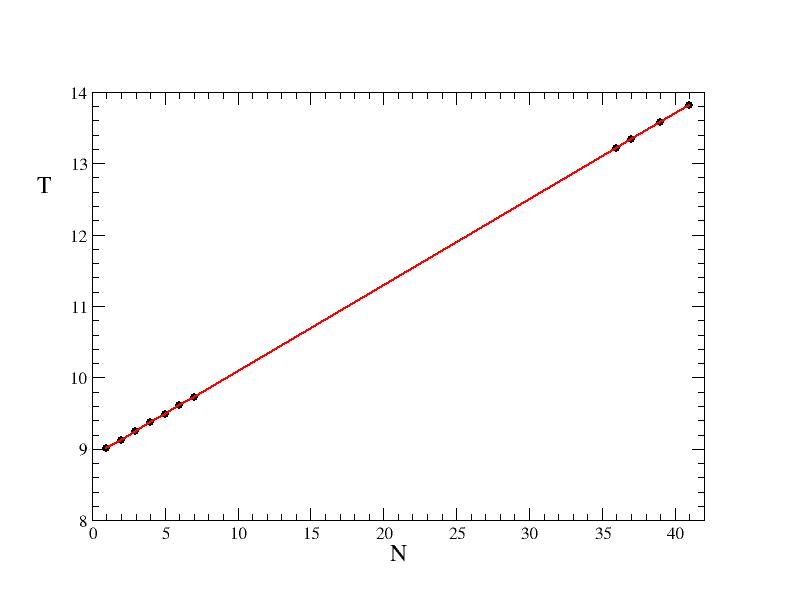}
	\caption{\label{fig:timebounce} The time $T$ as a function of the window index $N$ for $m=2, \alpha=1$.}
\end{figure}

It was found that the time `$T$' between the first and second collisions of the two-resonance windows is related to the $\omega_{R}$ mode by the relation 
\begin{equation}
	\omega_{R} T  = 2 \pi N + \delta, \label{eq:linear_relation}
\end{equation}
where $N$ is an integer and $\delta$ is a phase constant. The requirement that $\delta$ be between $0$ and $2\pi$ fixes the number $N$ assigned to the first two windows to be $1$ and $2$, as shown in figures \ref{fig:first_two-resonance} and \ref{fig:second_two-resonance}, respectively. Figure \ref{fig:timebounce} shows that the linear relation cf. Eq. \eqref{eq:linear_relation} fits the numerical results well, with the resonance parameters resulting from the fit being $\omega_{R} =0.7068$ and $\delta=0.08491$. Here, $\omega_R$ does not coincide with any of the observed frequencies of the kink's excitation modes, which are given as $\omega_1=2.826$, $\omega_2=2.479$, and $\omega_3=1.659$, respectively. Neither does it agree with the frequency modes of the kink-antikink system, which are respectively given as $\omega_{1}= 3.241$, $\omega_{2} = 2.604$, and $\omega_{3} = 1.898$. In computing these frequency modes of the kink-antikink system, we follow similar analyzes of the single kink in section \ref{sec:models}. Considering a small perturbation of the kink-antikink configuration $\varphi(t,x) = \varphi_{\bar{K}} \left(x-x_{0}\right) + \varphi_{K} \left(x+x_{0}\right)-1 + \delta \varphi(t,x)$, where we have taken $\gamma = \frac{1}{\sqrt{1-v_{i}^{2}}} =1 $ we get equation \eqref{eq:wave_equ}, with $\varphi_{K}$ being replaced by the configuration in equation \eqref{eq:spec}. Thus, the frequency in the kink-antikink system does not correlate much with $\omega_{R}$. This result conform with similar calculations of the $\varphi^{4}$ model~\cite{Lee:2016lhd}.
The small value of $\omega_{R}$ results in the large observed time intervals between the two collisions and hence makes it easy for the energy transference from translational to internal mode to be achieved. The large deviation of $\omega_{R}$ from the observed internal modes can be attributed to the distortion between the kink-antikink interaction~\cite{Takyi:2021jzx,Gani:2015cda}. In the following step, we examine the relationship between the binding energy `$\beta$' of the kink-antikink interaction and $T$, as expressed by the relation~\cite{Campbell:1983xu, Campbell:1986nu} 
\begin{equation}
	\beta = T \sqrt{v_{c}^{2}-v_{i}^2}
\end{equation} 
which together with Eq. \eqref{eq:linear_relation} are used to predict the locations of the resonance windows:
\begin{equation}
	v_{i}^{\mathrm{predict}} = \frac{\sqrt{v_{c}^{2}\left(2N\pi + \delta\right)^{2}-\beta^{2}\omega_{R}^{2}}}{2N\pi + \delta}.
\end{equation}
The $6$th column of Table \ref{t2} gives the predicted resonance centers. Apart from $N=1$ which predicts well with the numerical value, the others show a large deviation from the numerical values with an error of $8$\% to $28$\%. 
As before, the deviations in these results can be attributed to the distortions in the kink-antikink interactions.
\begin{table}
	\centerline{
		\begin{tabular}{|c| c| c| c| c| c|c|}
			$N$ &  Range of $v_{i}$  &  Center  &  $T$   &  $\beta$  &  Predicted Center & Error \cr
			\hline
			1   &  0.1540-0.1560     &  0.1550  & 9.009  &  1.2670   &  0.1550           &  0.0\% \cr 
			\hline 
			2   &  0.1561-0.1581     &  0.1571  & 9.129  &  1.2625   &  0.1970           & 25.4\% \cr 
			\hline 
			3   &  0.1582-0.1599     &  0.15905 & 9.249  &  1.2583   &  0.2039           & 28.2\% \cr 
			\hline 
			4   &  0.1600-0.1610     &  0.1605  & 9.369  &  1.2586   &  0.2063           & 28.5\% \cr 
			\hline
			5   &  0.1611-0.1625     &  0.1618  & 9.489  &  1.2598   &  0.2073           & 28.1\% \cr 
			\hline 
			6   &  0.1626-0.1640     &  0.1635  & 9.609  &  1.2556   &  0.2079           & 27.2\% \cr 
			\hline 
			7   &  0.1641-0.1649     &  0.1645  & 9.729  &  1.2590   &  0.2082           & 26.5\% \cr 
			\hline 
			36  &  0.1927-0.1929     &  0.1928  & 13.213 &  1.07624  &  0.209280         & 8.5\%  \cr
			\hline 
			37  &  0.1930-0.1934     &  0.1932  & 13.333 &  1.07331  &  0.209274         & 8.3\%  \cr 
			\hline 
			39  &  0.1936-0.1939     &  0.19375 & 13.573 &  1.07454  &  0.209277         & 8.0\%  \cr 
			\hline 
			41  &  0.1941-0.1946     &  0.1943  & 13.813 &  1.07475  &  0.209279         & 7.7\%  \cr
			\hline 
	\end{tabular}}
	\caption{\label{t2} Analysis of the two-resonance windows in the kink-antikink collision for $m=2$, $\alpha=1$. The error is the maximum relative error between the predicted and numerical values of the resonance center.}
\end{table}

\section{Conclusion}
\label{sec:conclude}
In the present work, we have studied kink-antikink collisions for the noncanonical nonintegrable $\varphi^6$ model in one space and one time dimension. We were particularly interested in investigating which values of the parameter $\alpha$, control the curvature of the potential yield localized inner structures for $m=1$ and $m=2$ in the energy density of the kink. We also looked into whether the presence of localized inner structures coupled with shape modes would result in the generation of oscillons and resonance structures.

Starting from a general total energy $E =\int \rho(x)\,\mathrm{d}x$ where $\rho(x)$ is the energy density, we explored the dynamical properties of this model. In the regime where $m=1,2$ and $\alpha >0$, we obtained the excitation spectrum with four bound states: a zero mode responsible for translation and three internal modes which are crucial for the resonance windows. Also in this regime, we observed two to three localized inner structures in the energy density of the kink.

In studying the collision of the kink and antikink, we first take the superposition of a kink $\varphi_{K} \left(x, x_{0}, v_{i}, t=0\right)$ and antikink $\varphi_{\bar{K}} \left(x, -x_{0}, v_{i}, t=0\right)$. We then solve the dynamical equation of motion using the fourth-order center difference scheme. We reported from our numerical results the production of two to three oscillons for some initial velocities in the regime of $m=1, 2$ and $\alpha>0$. We also reported several resonance windows in this regime, and an analysis of the resonance mechanism was carried out. We found from this analysis that the resonance frequency falls short when compared to the excitation modes of the single kink as well as the kink-antikink system of this model. Also, the theoretically predicted centers of the resonance windows resulting from this analysis deviate largely from the numerical results. The large deviations from this model are a result of distortions that occur during the interaction of the kink and antikink.

\acknowledgments
We are grateful to Prof. H. Weigel and Prof. V. A. Gani for reading the manuscript and for helpful comments.


\begin{thebibliography}{99}
\bibitem{Ablowitz:1979a} M.~J.~Ablowitz, M.~D.~Kruskal, and J.~F.~ Ladik, SIAM~J.~Appl.~Math. \textbf{36} (1979) 428.

\bibitem{Moshir:1981ja} M.~Moshir, Nucl. Phys. B \textbf{185} (1981) 318.

\bibitem{Campbell:1983xu} D.~K.~Campbell, J.~F.~Schonfeld, and C.~A.~Wingate, Physica D \textbf{9} (1983) 1. 

\bibitem{Belova:1985fg} T.~I.~Belova and A.~E.~Kudryavtsev, Physica D \textbf{32} (1988) 18.
	
\bibitem{Anninos:1991un} P.~Anninos, S.~Oliveira, and R.~A.~Matzner, Phys. Rev. D \textbf{44} (1991) 1147.

\bibitem{Goodman:2005ja} R.~H.~Goodman and R.~Haberman, SIAM~J.~Appl.~Math. \textbf{4} (2005) 1195.

\bibitem{Peyrard:1983rzn} M.~Peyrard  and D.~K.~Campbell, Physica D \textbf{9} (1983) 33.

\bibitem{Dorey:2011yw} P.~Dorey, K.~Mersh, T.~Romanczukiewicz, and Y.~Shnir, Phys.~Rev.~Lett \textbf{107} (2011) 091602.

\bibitem{Weigel:2013kwa} H.~Weigel, J.~Phys.~Conf.~Ser \textbf{482} (2014) 012045.

\bibitem{Gani:2014gxa} V.~A.~Gani, A.~E.~Kudryavtsev, and M.~A.~Lizunova, Phys.~Rev.~D \textbf{89} (2014) 125009.

\bibitem{Gani:2015cda} V.~A.~Gani, V.~Lensky, and M.~A.~Lizunova, JHEP \textbf{08} (2015) 147.

\bibitem{Marjaneh:2017mko} A.~M.~Marjaneh, V.~A.~ Gani, D.~Saadatmand, S.~V.~Dmitriev, and K.~Javidan, JHEP \textbf{07} (2017) 028.

\bibitem{Belendryasova:2017wad} E.~Belendryasova and V.~A.~Gani, Commun.~Nonlinear~Sci.~Numer.~Simul \textbf{67} (2019) 414.  

\bibitem{Christov:2018wsa} I.~C.~Christov, R.~J.~Decker, A.~Demirkaya, V.~A.~Gani, P.~G.~Kevrekidis, and R.~V.~Radomskiy, Phys. Rev. D \textbf{99} (2019) 016010. 

\bibitem{Manton:2018deu} N.~S.~Manton, J.~Phys.~A \textbf{52} (2019) 065401. 

\bibitem{Gani:2020wej} V.~A.~Gani and A.~M.~Marjaneh, J. Phys. Conf. Ser. \textbf{1690} (2020) 012096.

\bibitem{Gani:2020pio} V.~A.~Gani, A.~M.~Marjaneh, and P.~A.~Blinov, Phys.~Rev.~D \textbf{101} (2020) 125017. 

\bibitem{Takyi:2016tnc} I.~Takyi and H.~Weigel, Phys.~Rev.~D \textbf{94} (2016) 085008. 

\bibitem{Demirkaya:2017euk} A.~Demirkaya, R.~Decker, P.~G.~Kevrekidis, I.~C.~ Christov, and A.~Saxena, JHEP \textbf{12} (2017) 071

\bibitem{Christ:1975wt} N.~H.~Christ and T.~D.~Lee, Phys. Rev. D \textbf{12} (1975) 1606.

\bibitem{Lohe:1979mh} M.~A.~Lohe, Phys. Rev. D \textbf{20} (1979) 3120.

\bibitem{Christov:2008kk} I.~Christov and C.~I.~Christov, Phys.~Lett.~A \textbf{372} (2008) 841.

\bibitem{Manton:1978gf} N.~S.~Manton, Nucl.~Phys.~B \textbf{150} (1979) 397.

\bibitem{Manton:2021ipk} N.~S.~Manton, K.~Oles, T.~Romanczukiewicz, and A.~Wereszczynski, Phys.~Rev.~Lett. \textbf{127} (2021) 071601. 

\bibitem{Rajaraman:1982is} R. Rajaraman,\emph{ Solitons and Instantons} (North Holland, 1982).

\bibitem{Manton2004} N.~Manton and P.~Sutcliffe \emph{Topological Solitons} (Cambridge University Press, 2004).

\bibitem{Vachaspati:2006zz} T. Vachaspati, \emph{Kinks and domain walls: An introduction to classical and quantum solitons} (Cambridge University Press, 2010).

\bibitem{Vilenkin:2000jqa} A.~Vilenkin and E.~P.~S Shellard, \emph{Cosmic Strings and Other Topological Defects} (Cambridge University Press, 2000).

\bibitem{Ivanov:1992aa} B.~Ivanov, A.~Kichiziev, and Y.~N.~Mitsai, Sov. Phys. JETP \textbf{75} (1992) 329.

\bibitem{Trullinger:1976aa} A.~R.~Bishop, J.~A.~Krumhansl, and S.~E.~Trullinger J. Physica D \textbf{1} (1980) 1.

\bibitem{Kevre2008} P.~G.~Kevrekidis, J.~Dimitri J and R.~Carretero-Gonz{\'a}lez \emph{Emergent Nonlinear Phenomena in Bose-Einstein Condensates: Theory and Experiment} (Springer-Verlag, 2008).

\bibitem{Greenwood:2008qp} E.~Greenwood, E.~Halstead, R.~Poltis and D. Stojkovic, Phys. Rev. D \textbf{79} (2009) 103003.

\bibitem{Ahlqvist:2014uha} P.~Ahlqvist, K. Eckerle and B. Greene, JHEP \textbf{04} (2015) 059.

\bibitem{Weigel:2008zz} H.~Weigel, Lect. Notes Phys. \textbf{743} (2008) 1.

\bibitem{Weigel:2021pbr} H.~Weigel and I. Takyi, Symmetry \textbf{13} (2021) 108.

\bibitem{Chiba:1999ka} T.~Chiba, T.~Okabe and M.~Yamaguchi, Phys. Rev. D \textbf{62} (2000) 023511.

\bibitem{Gomes:2013bca} A.~R.~Gomes, R.~Menezes, K.~Z.~Nobrega, and F.~C.~Simas, Phys. Rev. D \textbf{90} (2014) 065022.

\bibitem{Zhong:2019fub} Y.~Zhong, X.~L.~Du, Z.~C.~Jiang, Y.~X.~Liu, and Y.~Q.~Wang, JHEP \textbf{02} (2020) 153.

\bibitem{Takyi:2021jzx} I.~Takyi, B.~Barnes, H.~M.~Tornyeviadzi, and J.~Ackora-Prah, Turk. J. Phys. \textbf{46} (2022) 37.

\bibitem{Bazeia:2008tj} D.~Bazeia, L.~Losano, and R.~Menezes, Phys. Lett. B \textbf{668} (2008) 246.

\bibitem{Zhong:2014kha} Y.~Zhong and Y.~X.~Liu, JHEP \textbf{10} (2014) 041.

\bibitem{Zhong:2018tbi} Y.~Zhong, R.~Z.~Guo, C.~E.~Fu, and Y.~X.~Liu, Phys. Lett. B \textbf{782} (2018) 346.

\bibitem{Campbell:1986nu} D.~K.~Campbell, M.~Peyrard and P.~Sodano, Physica D \textbf{19} (1986) 165.

\bibitem{Gani:2017yla} V.~A.~Gani, A.~M.~Marjaneh, A.~Askari, E.~Belendryasova and D.~Saadatmand, Eur. Phys. J. C \textbf{78} (2018) 345.

\bibitem{Gani:2019jzc} V.~A.~Gani, A.~M.~Marjaneh and D.~Saadatmand, Eur. Phys. J. C \textbf{79} (2019) 620.

\bibitem{Bazeia:2017rxo} D.~Bazeia, E.~Belendryasova and V.~A.~Gani, Eur. Phys. J. C \textbf{78} (2018) 340.

\bibitem{Bazeia:2019nsh} D.~Bazeia, R.~A.~ Gomes, K.~Z.~ Nobrega, C.~F.~Simas, Phys. Lett. B \textbf{803} (2020) 135291.

\bibitem{Campbell:1986mg} D.~K.~Campbell and M.~Peyrard, Physica D \textbf{18} (1986) 47.

\bibitem{Gani:1998jb} V.~A.~Gani and E.~A.~Kudryavtsev, Phys. Rev. E \textbf{60} (1999) 3305.

\bibitem{Lee:2016lhd} Z. Lee and H. Weigel, \emph{in 61st Annual Conference of the South African Institute of Physics} (2016) 512.


\end{thebibliography}
\end{document}